Polarization conversion of non-specular diffraction orders from metallic nanohole arrays studied by Fourier space polarimetry

Z.L. Cao[1)], K. Ding[2)], L.Y. Yiu[1)], Z.Q. Zhang[2)], C.T. Chan[2)], and H.C. Ong[1)]
[1)] Department of Physics, The Chinese University of Hong Kong, Shatin, Hong Kong, People's Republic of China
[2)] Department of Physics, The Hong Kong University of Science and Technology, Clear Water Bay, Hong Kong, People's Republic of China

We use Fourier space polarimetry to study the incident angle- and polarization-dependent rotation angle $\psi$ and ellipticity $\chi$ of different diffraction orders emerging from a two-dimensional periodic Au array. The array has square lattice and circular nanoholes and thus is achiral. We find no polarization conversion occurs if the diffraction orders lie in the incident plane. However, for the orders that are diffracted away from the incident plane, their $\psi$ and $\chi$ vary considerably with incident angle and polarization. In particular, dramatic changes in $\psi$ and $\chi$ are observed when Bloch-like surface plasmon polaritons (SPPs) are excited. The experimental results are consistent with the finite-difference time-domain simulations. The transverse spin carried by the SPPs and a discrete dipole model are used complementarily to elucidate such angular and polarization dependences.

# I. INTRODUCTION

Since the invention of diffraction gratings by David Rittenhouse and Joseph von Fraunhofer back in three centuries ago, periodic structures have flourished across the entire electromagnetic spectrum in different areas spanning from biology to astronomy [1]. One of the most successful usages of diffraction gratings is to disperse electromagnetic waves. The dispersion of polychromatic waves into different diffraction angles has placed gratings as a key component in monochromators [1], spectrometers [1], wavelength division multiplexors [2], ultrashort pulse compressors [3], signal modulators [4], frequency stabilizers [5], etc. However, after centuries, periodic structures have gone beyond simple wave discrimination to more fascinating light-matter interaction. When the length scale of the lattice and basis of the periodic structures is closer or even smaller than the wavelength, many new systems arise. Notable examples are photonic crystals [6], plasmonic structures [7], and metamaterials [8] that have aroused a spur of research and development worldwide. They manifest numerous phenomena such as photonic band gap [6], enhanced transmission [10], Dirac cone [11], negative refractive [12], epsilon near zero [13], topological effects [14], etc, and studies on them have been ongoing for more than decades. In particular, with the advancement of nanotechnology, optical structures can now be designed and fabricated at will to exhibit different linear and nonlinear behaviors [10,15,16]. More importantly, additional entities such as fluorescent dyes, quantum dots, and molecules can be introduced to the nanostructures, branching out to various applications from light emission to biosensing [17-19].

Nevertheless, when studying the optical properties of the plasmonic structures and metamaterials, it is important to understand the radiation damping process. For example, in plasmonic crystals where a flat metal surface is milled with subwavelength bases [20], how the Bloch-like surface plasmon polaritons (SPPs) decay radiatively govern the response of scattering [21], fluorescence [22], surface-enhanced Raman scattering [23], multiphoton emission [24], second harmonic generation [25], etc, that are currently of intense interest. In fact, the outgoing waves manifest themselves into different diffraction orders at specific angles with different intensities and polarizations. In other words, when scanning across the detection angle for a given wavelength, the polarization state of the radiation could change from linear to elliptical or vice versa continuously along with distinct intensity variation. Both the intensity and polarization properties need to be considered properly for understanding the underlying physics and in prior to the actual device implementation. In general, the conservations of translational and angular momenta are fulfilled when SPPs dissipate to free space. While the translational momentum governs the diffraction angles and intensities that have been studied frequently by the dispersion relations [26], little is known about the polarization, which arises profoundly from the angular momentum counterpart. Particularly, as the surface waves are known to possess transverse spin density due to their out-of-phase transverse and longitudinal electric field components, how their angular momenta affect the outgoing radiations deserves attention as both the spin and orbital angular momenta as well as their spin-orbit interaction could take part in the process [27]. Its knowledge is compelling evidently from several earlier works. SPPs have been reported to produce polarization conversion from conically mounted one-dimensional gratings for more than 20 years ago but the role of SPPs play in determining the polarization is still not yet understood [28]. Remarkably, a recently report by Maoz et al have observed, in spite of lacking intrinsic chirality, two-dimensional (2D) square lattice circular metallic hole arrays can produce much stronger circular dichroism than the gammadion metamaterials, illustrating the importance of SPPs on the extrinsic chirality [29]. Rodríguez-Fortuño et al have demonstrated the spin-orbit interaction by changing the incident circular polarization to directionally control the excitation of SPPs on a flat metal surface [30]. Under reciprocity, the same group has also shown the travelling SPPs can transfers their spin angular momentum (SAM) to define the polarization state of the outgoing radiation upon scattering from a nanoparticle [31]. All these examples clearly point out the fact that the angular momentum carried by SPPs has a profound influence on the resulting far-fields.

Therefore, studying the polarization states of the diffracted waves from the plasmonic structures and understanding their origin are of importance. We previously have studied the interplay between the nonresonant reflection background and the SPP radiation damping in the polarization conversion of the lowest specular order from metallic nanohole arrays [32]. For higher orders where multiple diffractions are present, a full characterization is essential. In addition, no connection between the angular momentum of SPPs and the polarization conversion has been made although it should be considered properly. In this work, we investigate the rotation angle $\psi$ and ellipticity $\chi$ of different diffraction orders from an achiral array by using incident angle-resolved polarimetry under a Fourier space microscope. The array has four-fold symmetry and does not exhibit any intrinsic chirality. When light is illuminated on the sample at different angles along the Γ-X direction, we find the diffraction orders that lie in the incident plane experience no polarization conversion upon p- and s-excitations, i.e. both $\psi$ and $\chi$ are zero. However, for the orders that are diffracted away from the incident plane, strong dependences of $\psi$ and $\chi$ on incident angle and polarization are observed. In particular, dramatic changes in $\psi$ and $\chi$ are seen when Bloch-like SPPs are excited. The angular profiles thus consist of the SPP resonances superimposed on the slowly varying nonresonant backgrounds. Our results agree with the finite-difference time-domain (FDTD) numerical simulations. We find the $\chi$ at SPP resonances are governed by the transverse SAM carried by the SPPs during the radiation damping process. On the other hand, the nonresonant $\psi$ and $\chi$ backgrounds can be explained qualitatively by calculating the transition probabilities to various diffraction channels based on an electric dipole model.

## II. EXPERIMENTAL METHODS

2D square lattice Au array is fabricated by using interference lithography as described previously [32]. The scanning electron microscopy (SEM) image of the sample is shown in Fig. 1(a), showing it has period P = 800 nm, hole radius R and depth H = 100 and 200 nm. As the thickness of the Au film is larger than the skin depth, the sample has no transmission. After preparation, the sample is placed on a computer controller goniometer for angle-resolved reflectivity spectroscopy. Specular reflectivity spectra taken under p- and s-excitations are measured as a function of incident angle $\theta_i$. By contour mapping the spectra with angle, we obtain the dispersion relations for mode identification [26].

To study the polarization properties of all diffractions, i.e. reflections, at the same time, we have constructed an incident angle- and polarization-resolved Fourier space microscope, as shown in Fig. 2(a). A HeNe laser at $\lambda$ = 633 nm exiting from a single mode fiber is first collimated by an achromatic lens and then passed through an incident polarizer before being focused onto the back focal plane of a magnification = 100X, numerical aperture NA = 0.9 objective lens [33]. The output beam thus evolves as a collimated beam incident on the sample at $\theta_i$ defined by $d = f \sin\theta_i$, where f is the focal length of the objective lens and d is the offset of the point source from the optical axis of the objective lens [34]. Therefore, by placing the entire illumination optics on a motorized translation stage, $\theta_i$ can be varied from 0° to 60° with angular resolution as small as 0.125°. The polarizer is oriented at a position so that the incident polarization is either parallel with or perpendicular to the incident plane, i.e. p- or s-incidence. The reflections from the sample are then collected by the same objective lens and are fed into an EMCCD camera via a Fourier lens system for Fourier space imaging. For determining $\psi$ and $\chi$, a quarter-wave plate and an analyzer can be selectively placed between the objective lens and the detection unit for measuring the four Stokes parameters $S_0$, $S_1$, $S_2$, and $S_3$. The parameters are related to the reflection intensities $I$ given as $S_0 = I(0°, 0°) + I(90°, 0°)$,

$$S_1 = I(0°,0°) - I(90°,0°) \quad , \quad S_2 = 2I(-45°,0°) - I(0°,0°) - I(90°,0°) \quad , \quad \text{and}$$

$S_3 = 2I(-45°,90°) - I(0°,0°) - I(90°,0°)$, where the parenthesis (α,β) defines the orientation of the analyzer and the phase retardation introduced by the quarter wave plate [35]. The transmission axis of analyzer can be set at α = 0°, -45°, and 90° with respect to the incident plane by either removing the quarter wave plate (i.e. β = 0°) or inserting the wave plate with the fast axis parallel to α = 0° (i.e. β = 90°) (see the inset of Fig. 2(a)) [35]. Therefore, the reflections at four different detection configurations, $I(0°,0°)$, $I(90°,0°)$, $I(-45°,0°)$, and $I(-45°,90°)$, enable one to determine all Stokes parameters. Finally, ψ and χ are defined as $tan\,2\psi = S_2/S_1$ and $sin\,2\chi = S_3/S_0$ [35].

## III. RESULTS
### a. Dispersion relations

First, the p- and s-polarized dispersion relations of the array taken along the Γ-X direction are shown in Fig. 1(b)&(c) for identifying the SPP modes. The dispersive reflectivity dips indicate the excitation of (-1,±1) and (1,0) propagating Bloch-like SPPs, which are illustrated by the dash lines as deduced from the phase-matching equation based on the empty lattice approximation [26,32]:

$$\left(\frac{2\pi}{\lambda}\right)^2 \frac{\varepsilon_{Au}}{\varepsilon_{Au}+1} = \left(\frac{2\pi}{\lambda}\sin\theta_i + \frac{2m\pi}{P}\right)^2 + \left(\frac{2n\pi}{P}\right)^2, \quad (1)$$

where $\varepsilon_{Au}$ is the dielectric constant of Au obtained from Ref [36] and (m,n) is Bragg scattering order. For λ = 633 nm, as indicated by the solid lines, one sees the (-1,±1) and (1,0) SPPs are excited at $\theta_i$ = 5.5° and 15.5°, respectively, under p-incidence. On the other hand, a (-1,±1) mode is excited at $\theta_i$ ~ 4.5° under s-polarization by simple extrapolation considering the lowest detection angle for our system is 5°. The presence of s-excited (-1,±1) mode is due to the coupling of two degenerate (-1,1) and (-1,-1) SPPs, yielding the (-1,±1)$_s$ and (-1,±1)$_a$ modes with distinctive field symmetries and radiation damping [37]. The (-1,±1)$_s$ mode is symmetric with respect to the incident plane and is relatively nonradiative while the (-1,±1)$_a$ mode is asymmetric and more radiative. As a result, the symmetric p-polarized light can couple to the dark (-1,±1)$_s$ mode whereas the bright (-1,±1)$_a$ mode is excited by the s-polarized light [37]. Finally, a small plasmonic band gap and two bright and dark modes due to the coherent coupling between the (-1,±1) and (1,0) SPPs are found at λ = 662 nm and $\theta_i$ ~ 12.5° under p-excitation [37].

### b. Fourier space polarimetry

Once the SPP modes have been identified, the sample is then transferred to a Fourier space microscope for polarimetric imaging. As an example, Fig 2(b) shows the p-incident $I(0°,0°)$ image of the array taken at $\theta_i$ = 5° in the Γ-X direction. The plane of incidence is indicated by the dash line. Five sharp reflection spots are observed revealing the diffraction orders and they can be deduced by using the 2D grating equation given as:

$$\{k_x, k_y\} = \left\{\frac{2\pi}{\lambda}\sin\theta_i + \frac{2p\pi}{P}, \frac{2q\pi}{P}\right\}, \quad (2)$$

where $k_x$ and $k_y$ are the in-plane wavevector components and {p,q} are the diffraction orders. Therefore, we identify the specular {0,0} order at $\{k_x,k_y\}$ = {0.86,0} μm$^{-1}$, the {-1,0} order at {-7,0} μm$^{-1}$, the {1,0} order at {8.7,0} μm$^{-1}$, and two {0,±1} orders at {0.86,±7.9} μm$^{-1}$. While the specular and the {±1,0} reflections lie in the incident plane along the $k_x$ direction, both {0,±1} orders are diffracted away from the plane. It is noted that at the incident angles corresponding to the SPP excitations, these diffractions actually define the radiation damping channels of the SPPs. Fig. 2(c)-(e) display the $I(45°,0°)$, $I(45°,90°)$, and $I(90°,0°)$ images, showing the variation of the diffraction intensities under different configurations. The images are then used for calculating the Stokes parameters.

For each image, we integrate the spots to obtain the intensities of the orders. The intensities of the {0,0}, {-1,0} and {0,±1} orders are plotted as a function of incident angle in the Supplementary Information for reference [38]. We estimate the diffraction power ratios between {0,±1} and {0,0} orders are ~ 0.1 and 0.15 for SPPs under p- and s-excitations [38]. We then determine their corresponding ψ and χ in Fig. 3(a)-(d). As shown in the inset of Fig. 3(e), for the polarization state, the ellipse is perpendicular to the propagation direction of the diffractions and ψ is defined with respect to the p-polarization axis. The figures reveal for the orders lying in the incident plane, i.e. {0,0} and {-1,0}, on average, almost zero ψ and χ are found indicating no polarization conversion occurs. In fact, considering our square lattice system with circular holes, it does not possess any intrinsic chirality and thus no conversion is expected. However, for the {0,±1} orders, noticeable ψ and χ are observed and they vary considerably with $\theta_i$ for two polarizations. For example, for the {0,1} order taken under the p-excitation, Fig. 3(a), ψ begins at 90° and then increases slightly with $\theta_i$, before changing abruptly at $\theta_i$ = 6° and 15° where the (-1,±1)$_s$ and (1,0) SPPs are excited, exhibiting two asymmetric lineshape profiles. On the other hand, χ in Fig. 3(b) has a slowly varying positive background but is superimposed with two strong peak and dip at the excitation of the (-1,±1)$_s$ and (1,0) SPPs. For the s-incidence in Fig. 3(c)&(d), ψ increases from zero gradually with increasing $\theta_i$ and display a similar asymmetric profile at the (-1,±1)$_a$ SPPs. At the same time, χ increases with $\theta_i$ and peaks at the (-1,±1)$_a$ SPPs but then decreases to negative afterwards in Fig. 3(d). For the {0,-1} order, it undergoes almost the same polarization change as the {0,1} order but in opposite sign. In other words, the polarization properties of the {0,±1} orders possess a mirror image, which is expected from our mirror symmetric system.

We visualize the {0,1} order under p- and s-polarizations in Fig. 3(e)&(f) to have a better physical picture. Under normal p-incidence, the order is linearly polarized but oriented perpendicular to the diffraction plane. Therefore, it is s-polarized even under a p-incidence. However, when $\theta_i$ increases, the order becomes a left elliptically polarized light with its major axis tilted towards the p-axis. At the excitation of the (-1,±1)$_s$ SPPs, although the order remains left elliptically polarized, it becomes more circular and the major axis is s-oriented. In addition, due to the asymmetric profile, the major axis of the ellipse swings between the s-axis. After that, the major axis slowly moves away from the s-axis but swings again at the excitation of the (1,0) SPPs. In addition, it switches to right elliptically polarized since χ now is negative. Likewise, under the s-incidence, we see at $\theta_i$ = 0° the order is now a p-polarized light, showing the polarization of the out-of-plane order under normal incidence is always perpendicular to that of incidence regardless of the excitation polarization. When $\theta_i$ increases, it becomes left elliptically polarized with its major axis tilted away from the p-axis. Around the (-1,±1)$_a$ SPPs, the order reproduces the behavior of the (-1,±1)$_s$ SPPs but becomes right elliptically polarized at larger angle.

c. **Finite-difference time-domain simulations**

To confirm our experimental result, we have conduct FDTD simulations and the unit cell is shown in Fig. 1(a). It has period, hole radius, and depth = 800, 120, and 100 nm. A small sinusoidal modulation with height = 30 nm is added with reference to the SEM image. Bloch boundary condition is used at four sides and perfectly match layer is set at the top and bottom of the cell. A power monitor is placed at 5 nm above the metal surface for calculating the diffractions. The dielectric constant from Ref [36] is used for Au. First, the angle-dependent specular p- and s-polarized reflectivity mappings for the array are shown in Fig. 1(d)&(e) calculated along the Γ-X direction. Good agreement between the simulation and experiment is seen. The $(-1,\pm1)_s$, $(1,0)$, and $(-1,\pm1)_a$ SPP modes are clearly seen from the simulations and they are excited at 5°, 15.5°, and 4.25° for λ = 633 nm. The plasmonic gap and two coupled modes are also reproduced well. We then calculate the ψ and χ as a function of $θ_i$ for four diffraction orders and the results are shown in Fig. 4(a)-(d) under two polarizations. We find the simulation and experiment fairly agree with each other although discrepancies exist, particularly the magnitudes of ψ and χ, mostly due to sample imperfections and optical misalignment. Nevertheless, the FDTD results reproduce most of the major features in experiment.

## IV. DISCUSSION

We attempt to explain the angle and polarization dependences of ψ and χ. The SPP mediated χ is first studied by examining the transverse SAM of SPPs. Then, an electric dipole mode is used to depict the ψ and χ backgrounds.

### a. SPP resonances

As χ is defined by the SAM, we examine the SAM carried by the SPPs and the $\{0,\pm1\}$ diffraction orders to search for any connection In general, the spin density of the plane wave is given as [39,40]:

$$\vec{s} = \text{Im}\left[\varepsilon_o \vec{E}^* \times \vec{E} + \mu_o \vec{H}^* \times \vec{H}\right]/4\omega, \quad (3)$$

where $\vec{E}$ and $\vec{H}$ are the electric and magnetic fields and $ε_o$ and $μ_o$ are the permittivity and permeability. The FDTD simulated SAM of two $\{0,\pm1\}$ diffraction orders calculated by Eq. (3) are shown in Fig. 5(a)&(b) and they resemble to χ in Fig. 3(b)&(d) and 4(b)&(d). In fact, for the SPP resonances, we speculate the transverse spin plays a major role in governing the polarization of the $\{0,\pm1\}$ diffractions. When the SPPs decay, the radiation damping process is subjected to the momentum conservations in which the translational and angular momenta of the SPPs will determine the outgoing wavevectors, i.e. the diffraction angles, and the polarization states . While the conservation of translational momentum gives rise to the phase matching equation in Eq. (1), the angular counterpart is not trivial as it relies on the spin and orbital angular momenta. However, for plane waves where no Goos-Hanchen and/or Imbert-Fedorov shifts are present [27], the change of orbital angular momentum is negligible throughout the damping process. Therefore, we conclude the total SAM of the diffractions ($\vec{s}_{diff}$) should be offset by that of SPPs ($\vec{s}_{SPP}$). For our case, since the specular and $\{\pm1,0\}$ reflections are all linearly polarized and do not carry SAM, $\vec{s}_{diff}$ is dictated primarily by $\vec{s}_{\{0,1\}} + \vec{s}_{\{0,-1\}}$. In addition, knowing both $\vec{s}_{\{0,1\}}$ and $\vec{s}_{\{0,-1\}}$ are equal in magnitude but opposite in direction, only the y-component of $\vec{s}_{\{0,1\}} + \vec{s}_{\{0,-1\}}$ i.e. $\left(\vec{s}_{\{0,1\}} + \vec{s}_{\{0,-1\}}\right) \cdot \hat{y}$, prevails as the x- and z-components will cancel out each other.

We examine the $(-1,\pm1)_{s,a}$ SPP modes by studying their near-field SAM. Given the SPP wavevectors qualitatively as $\vec{k}_{SPP} = (2\pi \sin\theta/\lambda - 2\pi/P)\hat{x} \pm 2\pi/P \,\hat{y} = -K_x\hat{x} \pm K_y\hat{y}$ from Eq. (1), the magnetic fields for two $(-1,\pm1)_{s,a}$ SPP standing waves can be approximated as: $2H_o e^{-iK_x x} e^{-K_z z}\left(-iK_y \sin(K_y y)\hat{x} - K_x \cos(K_y y)\hat{y}\right)$ and $2H_o e^{-iK_x x} e^{-K_z z}\left(-K_y \cos(K_y y)\hat{x} - iK_x \sin(K_y y)\hat{y}\right)$, where $H_o$ is a constant and $K_z$ is the penetration depth [41]. By using $\vec{E} = i(\nabla \times \vec{H})/\omega\varepsilon$, their corresponding electric fields are:

$$\frac{2iH_o}{\omega\varepsilon} e^{-iK_x x} e^{-K_z z}\left(-K_x K_z \cos(K_y y)\hat{x} + iK_y K_z \sin(K_y y)\hat{y} + i(K_x^2 + K_y^2)\cos(K_y y)\hat{z}\right) \quad \text{and}$$

$$\frac{2iH_o}{\omega\varepsilon}e^{-iK_x x}e^{-K_z z}\left(-iK_x K_z \sin(K_y y)\hat{x} + K_y K_z \cos(K_y y)\hat{y} + (-K_x^2 - K_y^2)\sin(K_y y)\hat{z}\right).$$ Clearly, from the expressions, we notice some electric and magnetic components are π/2 out of phase, resulting in spinning electric and magnetic fields in different directions [40]. The magnetic field carries SAM in the z-direction whereas the electric field has SAM in both the y- and z-directions. However, one can show by Eq. (3) that only the y-component survives as it exhibits $K_x \cos^2(K_y y)$ and $K_x \sin^2(K_y y)$ dependences for the symmetric and asymmetric SPPs but the x- and z-components follow $\cos(K_y y)\sin(K_y y)$ that will be neutralized to zero after integrating over all space [42]. Therefore, the $(-1,\pm1)_{s,a}$ modes, which propagate in the negative x-direction, exhibit transverse SAM in the positive y-direction, giving rise to the spin-momentum locking [43]. For verification, we use FDTD to calculate the SAM patterns for the $(-1,\pm1)_{s,a}$ modes at z = 10 nm for different directions and the results are shown in Fig. 6(a)–(f). In fact, consistent with the analytical predictions, we see the y-components for two modes display symmetric and asymmetric patterns, which agree with the $\cos^2(K_y y)$ and $\sin^2(K_y y)$ dependences. While the y-component is predominately positive across the unit cell, the alternating positive and negative SAM for other two components results in zero SAM after integrating over the whole region. On the other hand, for the far-field, as the {0,1} and {0,-1} diffraction orders are left- and right elliptically polarized, the vector sum of the SAM, $\vec{s}_{diff}$, is also pointing in the positive y-direction, revealing the SAM of SPPs actually dominates that of the diffractions.

The same argument applies to the (1,0) SPP mode. With $\vec{k}_{SPP} = (2\pi \sin\theta/\lambda + 2\pi/P)\hat{x} = K_x \hat{x}$, the electric field of (1,0) SPPs can be expressed as $E_0 e^{iK_x x}e^{-K_z z}(\hat{x} + iK_x/K_z \hat{z})$. The spin-momentum locking manifests both the SAM of the positive x-traveling SPPs and the vector sum of the {0,±1} orders pointing to the negative y-direction. The FDTD simulated SPP SAM patterns are shown in Fig. 6(g)–(i), which again are consistent with the analytical interpretations. We find the SAM is always negative in y-direction across the whole cell but is equally positive and negative for the x- and z-directions.

More insight can be provided by comparing the near- and far-field SAM as a function of incident angle. For each angle, we calculate by FDTD the near-field SAM integrated over the whole unit cell and the vector sum of the {0,±1} diffraction orders in Fig. 5(c)&(d) under p- and s-incidences. It is noted that the results in the figures are completely provided by the y-component and the x- and z-components for all angles are negligibly small. Therefore, both the near- and far-field SAM are always pointing towards the y-direction regardless of whether SPPs are excited or not. Two points are noted. First, the magnitude of the near-field SAM is much larger than that of the far-field. Such difference is reasonable as the near-field strength is usually much stronger than the far-field, particularly for SPP excitations. Second, the near-field SAM does not contribute to the nonresonant far-field background. As we see from Fig. 5(c)&(d), at large incident angles under p- and s-incidences, the near- and far-field SAM have different signs. All these imply other not-so-obvious angular momenta from the system should counteract the transverse spin of the near-fields to yield the SAM of the diffractions.

b. **Nonresonant ψ and χ backgrounds**

Analytically, we find the nonresonant backgrounds can be qualitatively explained within the framework of a discrete dipole model. For a square lattice array, under the condition where R << P and λ, each hole can be modeled as an electric dipole [42]. The transition probabilities to various diffraction channels are calculated, as given in the Supplementary Information [38]. The key element of the model is the optical transition matrix **T**, which is derived from the density matrix operator. It is noted that the **T**-matrix has already included the contributions from both the air-hole resonance as well as the lattice effect. It is expressed as [44]:

$$\mathbf{T} = \begin{pmatrix} T_{\parallel} & T_{xy} & \\ T_{xy} & T_{\parallel} & \\ & & T_{\perp} \end{pmatrix}, \quad (4)$$

where the explicit forms of the matrix elements are given in the Supplementary Information [38]. Once the matrix is ready, the transition amplitude from the incidence $\sigma_i$ to the diffraction $\sigma_d$, which is defined as $t_{\sigma_i \to \sigma_d} = \langle \hat{e}_{\sigma_d} | \mathbf{T} | \hat{e}_{\sigma_i} \rangle$, can then be formulated. The channel label is denoted as $\sigma_{i,d} = (\theta_{i,d}, \phi_{i,d}, s/p)$, where the first two angles stand for the incident and diffraction polar $\theta$ and azimuthal $\phi$ angles and the last quantity represents the excitation polarization, i.e. p- or s-polarization. As a result, the transition amplitudes from the p-polarized incidence to the s- and p-polarized {0,±1} diffraction orders can be expressed as:

$$t_{(\theta_i,0,p) \to (\theta_d,\phi_d,s)} = T_{\parallel} \sin\phi_d \cos\theta_i - T_{xy} \cos\phi_d \cos\theta_i$$
$$t_{(\theta_i,0,p) \to (\theta_d,\phi_d,p)} = T_{\parallel} \cos\theta_i \cos\theta_d \cos\phi_d + T_{xy} \cos\theta_i \cos\theta_d \sin\phi_d + T_{\perp} \sin\theta_i \sin\theta_d \quad (5)$$

where $\phi_i$ has been set at 0° due to the incidence is always along the Γ-X direction. Likewise, the amplitude for the s-polarized incident counterpart is obtained as:

$$t_{(\theta_i,0,s) \to (\theta_d,\phi_d,s)} = -T_{xy} \sin\phi_d + T_{\parallel} \cos\phi_d$$
$$t_{(\theta_i,0,s) \to (\theta_o,\phi_o,p)} = -T_{xy} \cos\theta_d \cos\phi_d - T_{\parallel} \cos\theta_d \sin\phi_d \quad (6)$$

In fact, Eq. (5) & (6) can be further simplified. When under normal incidence, we have $t_{(\theta_i,0,p) \to (\theta_d,\phi_d,s)} = \pm T_{\parallel}$ and $t_{(\theta_i,0,p) \to (\theta_d,\phi_d,p)} = \pm T_{xy} \cos\theta_d$ as well as $t_{(\theta_i,0,s) \to (\theta_d,\phi_d,s)} = \pm T_{xy}$ and $t_{(\theta_i,0,s) \to (\theta_o,\phi_o,p)} = \pm T_{\parallel} \cos\theta_d$ by setting $\theta_i = 0°$ and $\phi_d = \pm 90°$ for p- and s-excitations. Knowing from symmetry and the experimental and FDTD results that $\psi = \pm 90°$ and 0° but $\chi = 0°$ for two polarizations, we conclude $T_{\parallel} \gg T_{xy}$. Therefore, $T_{xy}$ can simply be ignored. Unfortunately, the precise evaluation of both $T_{\parallel}$ and $T_{\perp}$ is difficult and beyond the scope of this study.

We attempt to learn more of our case from the transition amplitudes. Both $\theta_d$ and $\phi_d$ can be deduced from the grating equation considering P and λ = 800 and 633 nm. For the {0,±1} orders, the grating equation is given as $\sin\theta_i \hat{x} \pm \frac{\lambda}{P} \hat{y} = \sin\theta_d (\cos\phi_d \hat{x} + \sin\phi_d \hat{y})$. Therefore, $\tan\phi_d = \pm 0.79/\sin\theta_i$ and $\sin\theta_d = \pm 0.79/\sin\phi_d$, which show $\phi_d$ decreases from 90° to ~ 62° but $\theta_d$ increases from 53° to 65° when $\theta_i$ increase from 0° to 25°. As a result, we expect, for p-excitation, the transition from p to s decreases while at the same time the p to p transition increases. On the other hand, for s-excitation, the s to s transition increases but that of the s to p decreases. Both are consistent with the ψ backgrounds. However, as $T_{xy}$, $T_{\parallel}$, and $T_{\perp}$ are not available, we only peek into χ by examining the relative phase of the transitions under s-incidence. $\frac{t_{(\theta_i,0,s) \to (\theta_d,\phi_d,s)}}{t_{(\theta_i,0,s) \to (\theta_o,\phi_o,p)}} = \frac{-1}{\cos\theta_d \tan\phi_d}$, implying the {0,±1} diffractions are plane wave-like and χ is small and almost constant with $\theta_i$ when compared with that of the p-counterpart. This agrees with our results. In addition, for positive $\phi_d$, i.e. the {0,1} diffraction, the s-wave lags behind the p-wave but leads ahead if $\phi_d$ is negative, i.e. the {0,-1} diffraction, suggesting the {0,1} order is right polarized whereas the {0,-1} is left. More importantly, $t_{(\theta_i,0,p) \to (\theta_d,\phi_d,s)} = -t_{(\theta_i,0,p) \to (\theta_d,-\phi_d,s)}$ and $t_{(\theta_i,0,p) \to (\theta_d,\phi_d,p)} = t_{(\theta_i,0,p) \to (\theta_d,-\phi_d,p)}$ for p-incidence and $t_{(\theta_i,0,s)\to(\theta_d,\phi_d,s)} = t_{(\theta_i,0,s)\to(\theta_d,-\phi_d,s)}$ and $t_{(\theta_i,0,s)\to(\theta_d,\phi_d,p)} = -t_{(\theta_i,0,s)\to(\theta_d,-\phi_d,p)}$, they imply the Stokes parameters $S_1^{\{0,+1\}} = S_1^{\{0,-1\}}$, $S_2^{\{0,+1\}} = -S_2^{\{0,-1\}}$ and $S_3^{\{0,+1\}} = -S_3^{\{0,-1\}}$, where the superscripts indicate the (0,±1) diffraction orders. Hence, both ψ and χ for {0,±1} diffraction orders should have the same magnitude but opposite in sign when varying $\theta_I$, supporting our results as well.

## V. CONCLUSION

In summary, we use Fourier space polarimetry to study the dependence of the rotation angle ψ and ellipticity χ of the diffraction orders arising from an achiral Au periodic array on incident angle. Along the Γ-X direction, we observe the specular and {-1,0} reflections, both lying in the incident plane, experience no polarization conversion. However, for the {0,±1} orders that are diffracted away from the incident plane, their ψ and χ vary considerably with angle under both p- and s-excitations. In particular, dramatic changes in ψ and χ occur when the Bloch-like (-1,±1) and (1,0) SPPs are excited. The electrodynamic simulations support the experimental results. We explain the angular profiles based on the transverse spin carried by the SPPs, which governs the total SAM of the out-of-plane diffracted orders, and a discrete dipole model.

## VI. ACKNOWLEDGEMENT

This research was supported by the Chinese University of Hong Kong through the Direct Grants 4053077 and 4441179, RGC Competitive Earmarked Research Grants, 402812 and 14304314, and Area of Excellence AoE/P-02/12.

**Figure Captions:**

1. (a) The plane-view scanning electron microscopy image of the Au nanohole array. The unit cell for the FDTD simulation. The experimental (b) p- and (c) s-polarized incident angle-dependent specular reflectivity mappings taken along the Γ-X direction. The dash lines are deduced from the phase matching equation, indicating (-1,±1)$_s$, (1,0), and (-1,±1)$_a$ SPP modes are excited. The solid line shows the λ = 633 nm. The FDTD simulated (d) p- and (e) s-polarized incident angle-dependent specular reflectivity mappings.

2. (a) The schematic of the incident angle resolved Fourier space polarimetry microscope. The layout for measuring the four Stokes parameters of the diffraction. $\vec{e}_p$ and $\vec{e}_s$ are defined as the p- and s-polarizations. (b)-(e) The Fourier space $I(0°,0°)$, $I(-45°,0°)$, $I(-45°,90°)$, and $I(90°,0°)$ images of the array taken at θ = 5º under p-incidence along the Γ-X direction. The specular {0,0} and {±1,0} reflection orders lie in the incident plane, as indicated by the dash line. The {0,±1} orders are diffracted away from the incident plane. The {0,0} and {0,±1} orders always lie in a constant $k_x$ line regardless of the incident angle. The {1,0} will move beyond the numerical aperture of the objective lens when $θ_i$ increases further.

3. The variations of the experimental (a,b) p- and (c,d) s-excited angle of rotation ψ and ellipticity χ of the specular {0,0}, {-1,0}, and {0,±1} diffraction orders as a function of incident angle. The excitation of the (-1,±1)$_s$, (1,0), and (-1,±1)$_a$ SPP modes are indicted by arrows. The visualizations of the polarization states of the {0,1} diffraction order under (e) p- and (f) s-incidence. The polarization ellipse is defined normal to the propagation direction of the diffraction and the ψ and χ are defined accordingly.

4. The dependences of the FDTD simulated (a,b) p- and (c,d) s-excited angle of rotation ψ and ellipticity χ of the specular {0,0}, {-1,0}, and {0,±1} diffraction orders on incident angle.

5. The FDTD simulated (a) p- and (b) s-excited SAM of {0, ±1} orders as a function of incident angle. The comparison between the SPP SAM and the vector sum of the {0,±1} diffraction orders under (c) p- and (d) s-incidences. Note the magnitudes of SPP SAM are much larger than those of the vector sum.

6. The FDTD simulated x-, y- and z-components ($s_x$, $s_y$, and $s_z$) of SAM for (a-c) (-1,±1)$_s$ mode, (d-e) (-1,±1)$_a$, and (g-i) (1,0) SPP modes in the unit cell. The scale bars are $10^{-27}$ J·s/m$^3$. The circles are the holes.

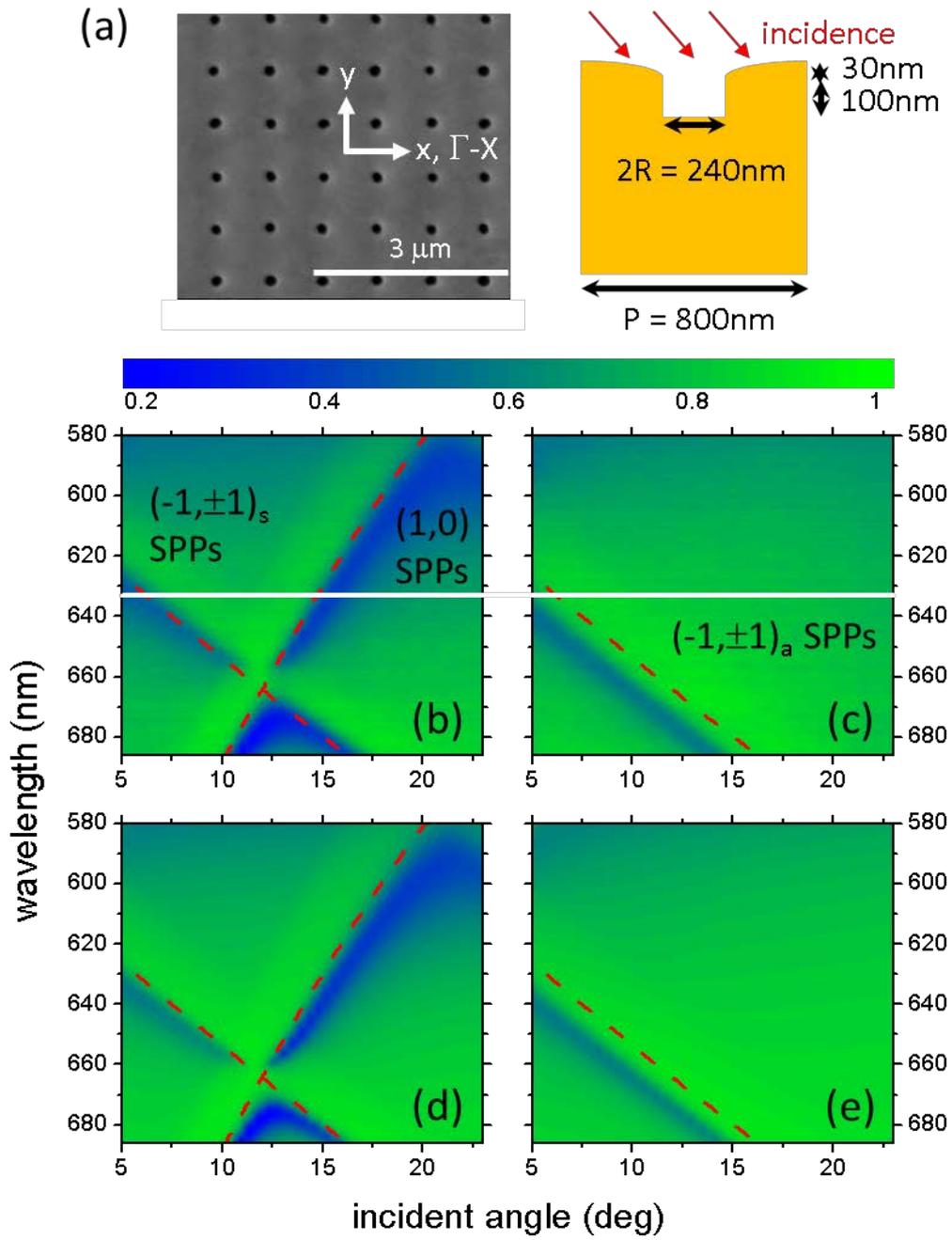

Fig. 1

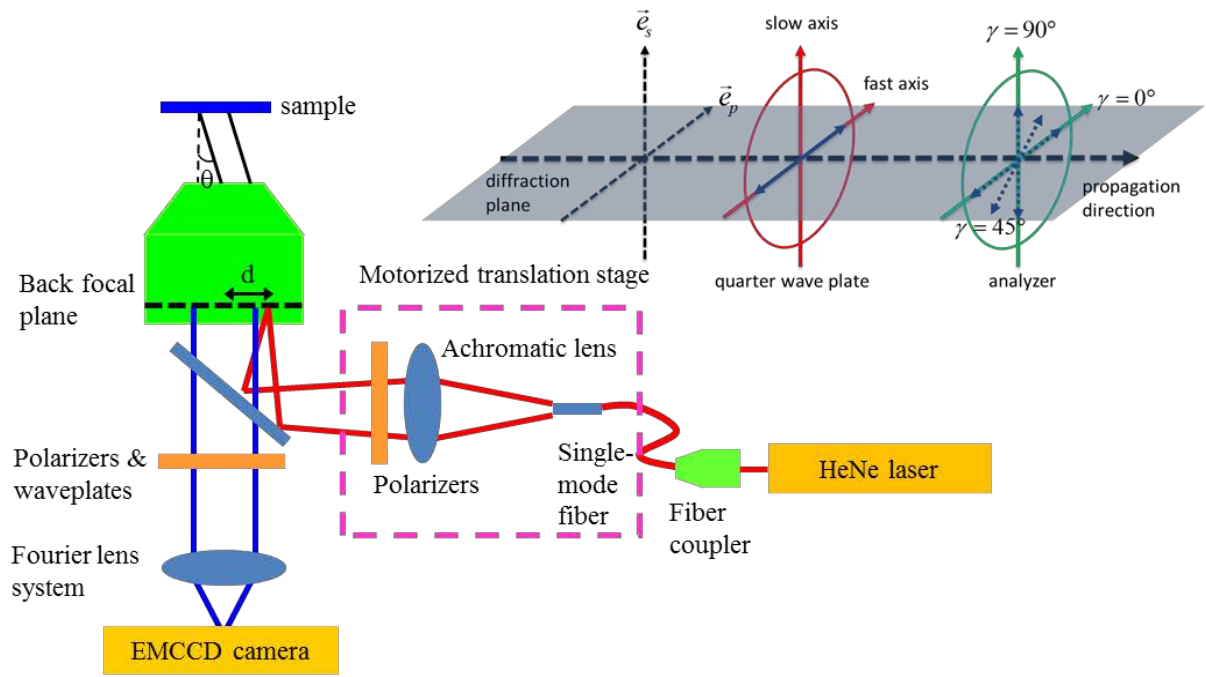
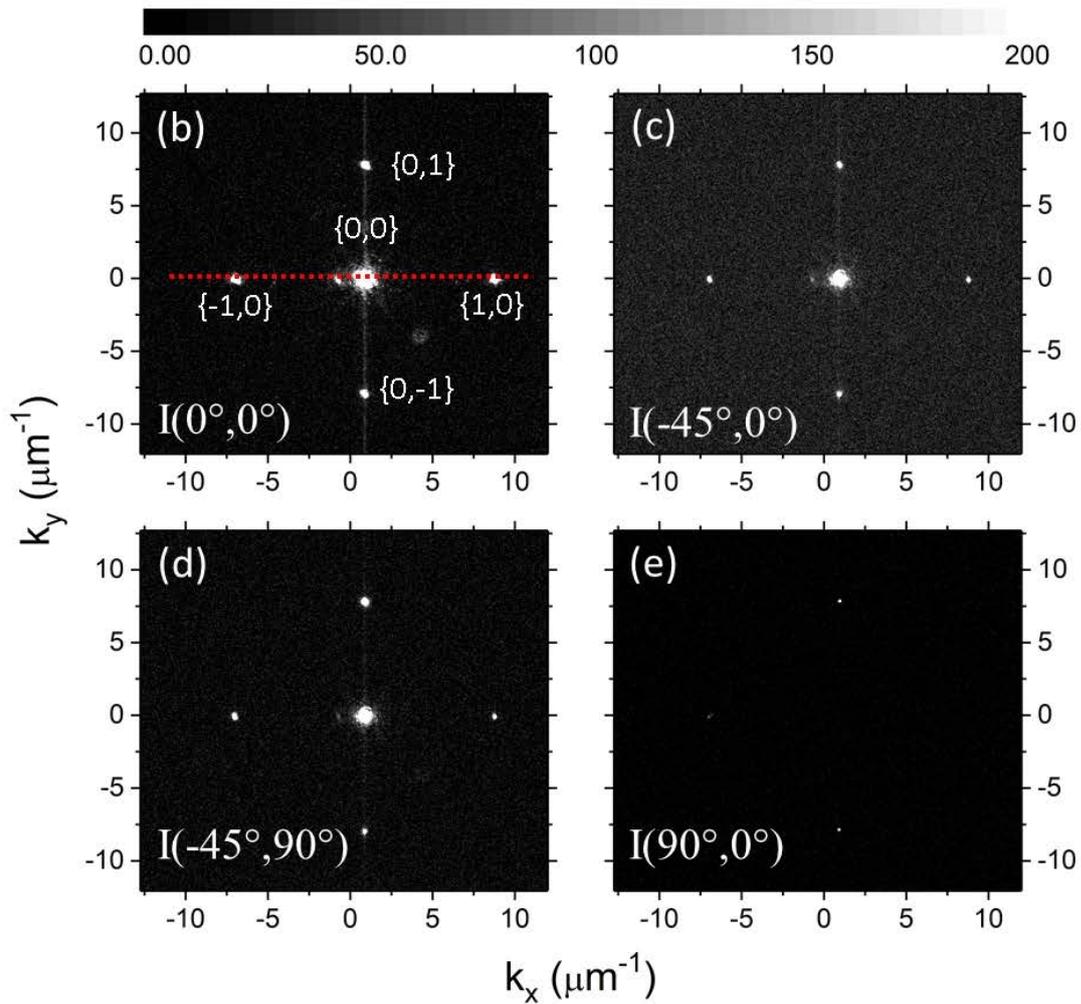

Fig. 2

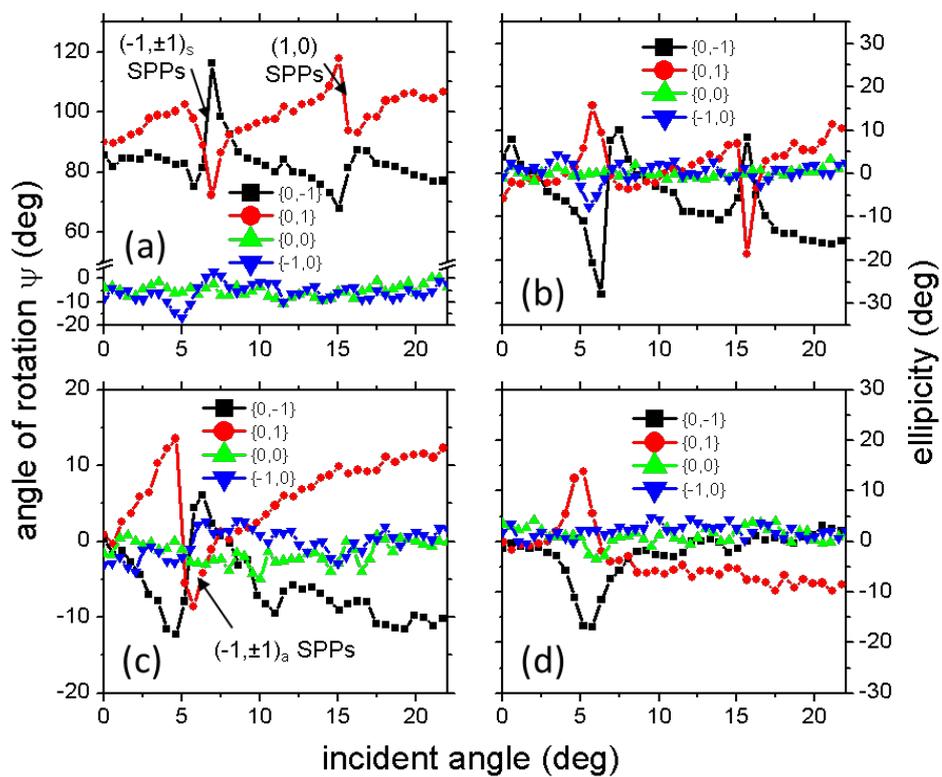
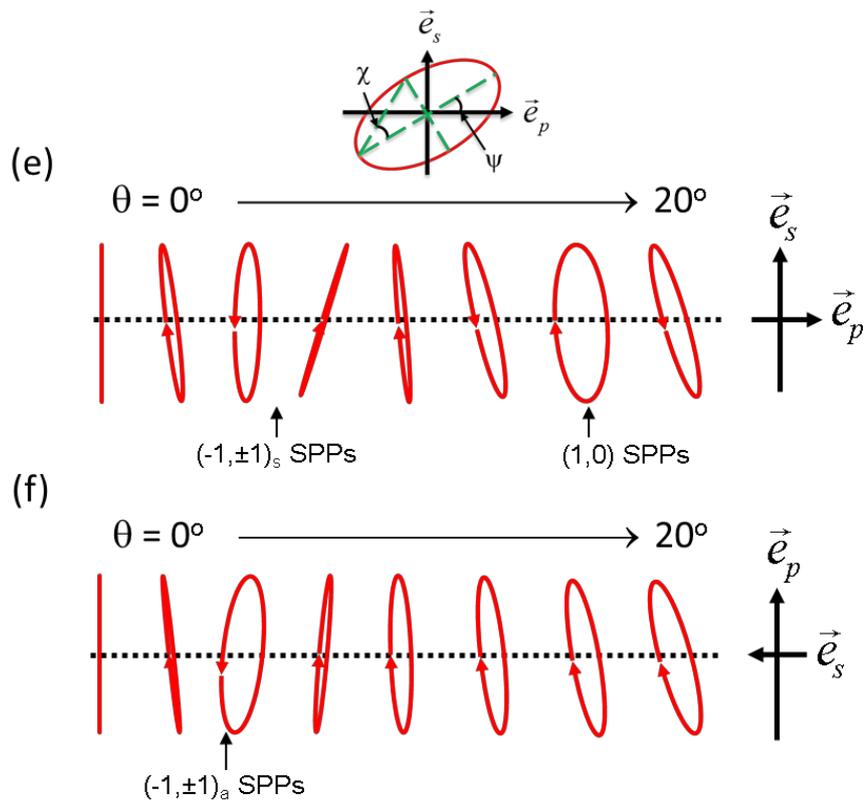

Fig. 3

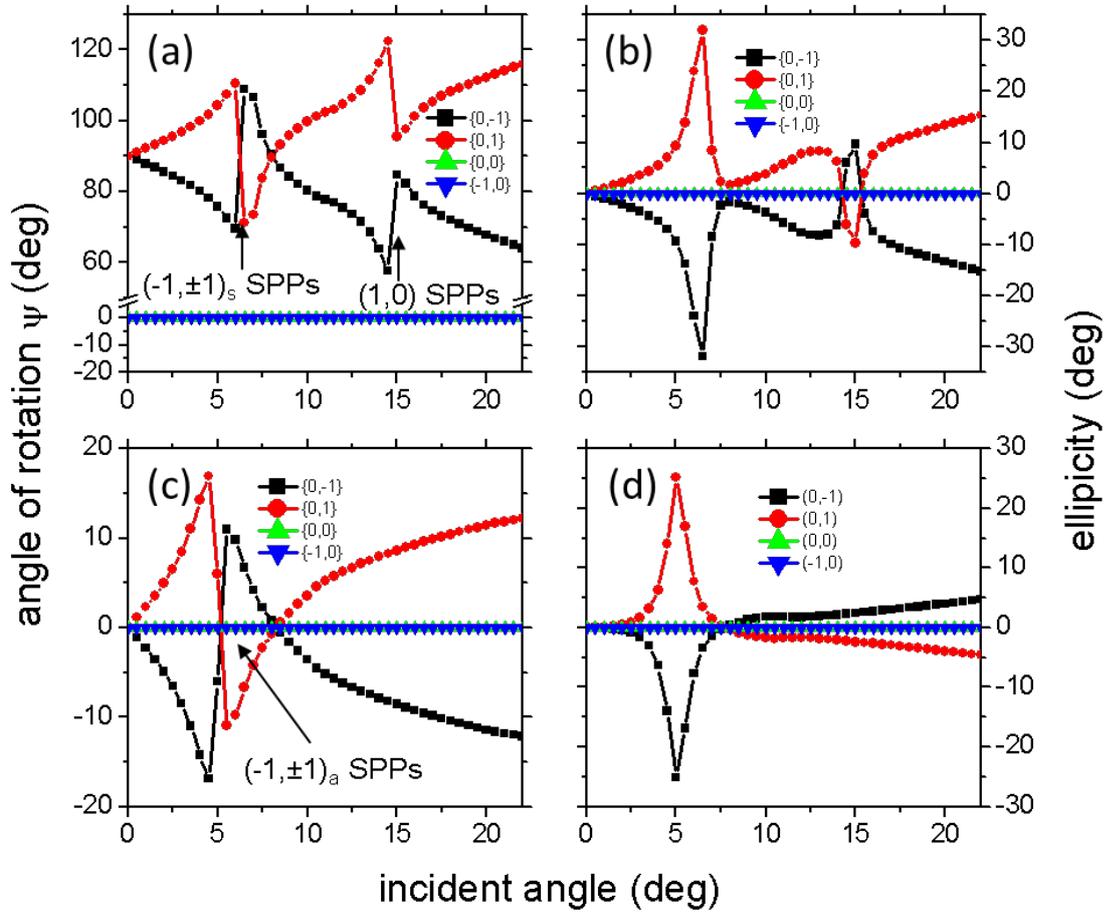

Fig. 4

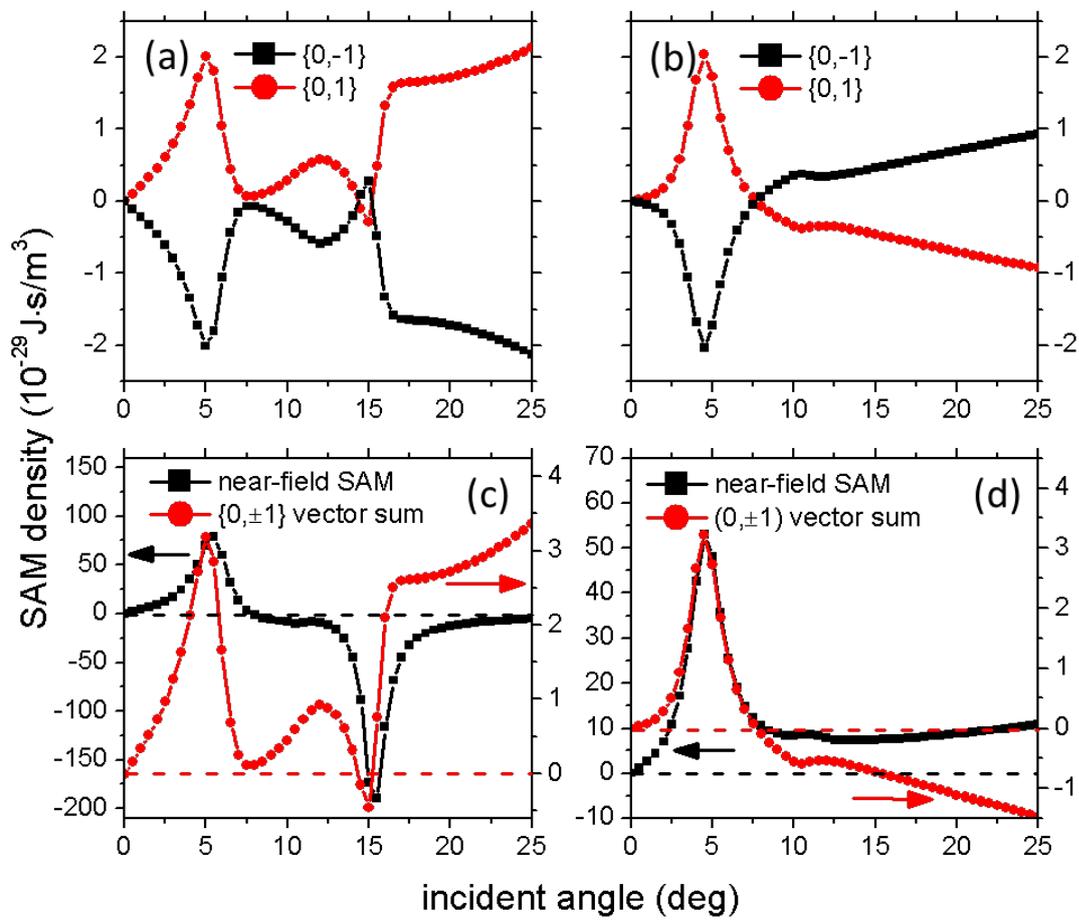

Fig. 5

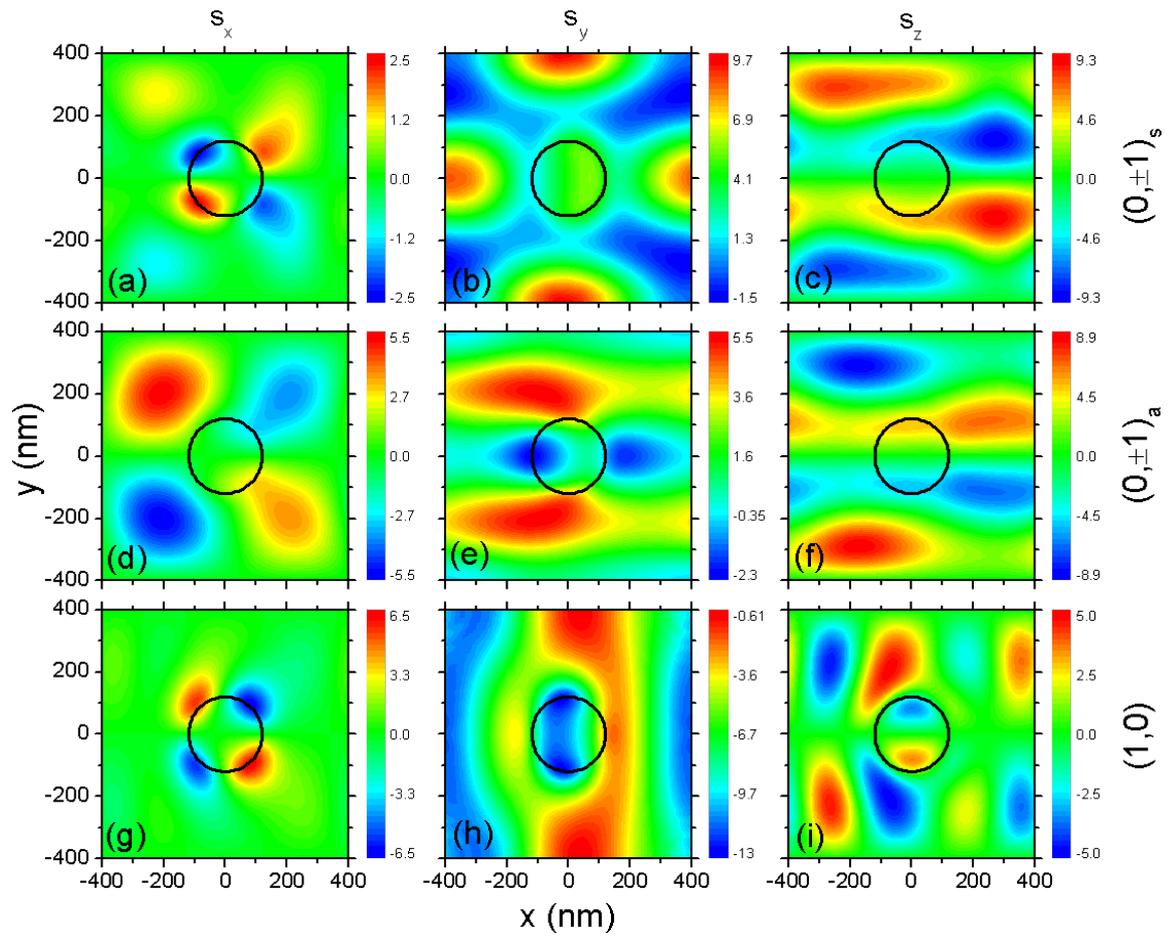

Fig. 6

# Supplemental Information for "Polarization conversion of non-specular diffraction orders from metallic nanohole arrays studied by Fourier space polarimetry"

Z.L. Cao[1], K. Ding[2], L.Y. Yiu[1], Z.Q. Zhang[2], C.T. Chan[2], and H.C. Ong[1]

[1] Department of Physics, The Chinese University of Hong Kong, Shatin, Hong Kong, People's Republic of China

[2] Department of Physics, The University of Science and Technology, Clear Water Bay, Hong Kong, People's Republic of China

## I. Plots of diffraction intensity with incident angle

The following figures show the plots of (a,e) $I(0°,0°)$, (b,f) $I(90°,0°)$, (c,g) $I(-45°,0°)$, and (d,h) $I(-45°,90°)$ of the {0,0}, {-1,0}, and {0,±1} diffraction orders as a function of incident angle under (a-d) p- and (e-h) s-excitations.

The diffraction power ratio between {0,±1} and {0,0} orders is defined as:

$$\frac{\text{diffracted powers from} \{0,\pm 1\} \text{ orders}}{\text{diffracted powers from} \{0,0\} \text{ order}} = \frac{\left[I(0^o,0^o)+I(90^o,0^o)\right]_{\{0,1\}} + \left[I(0^o,0^o)+I(90^o,0^o)\right]_{\{0,-1\}}}{\left[I(0^o,0^o)+I(90^o,0^o)\right]_{\{0,0\}}}$$

where the subscripts represent the diffraction orders. Therefore, from the plots, the power ratios are determined to be ~ 0.1 and 0.15 for SPP excitations under p- and s-incidences.

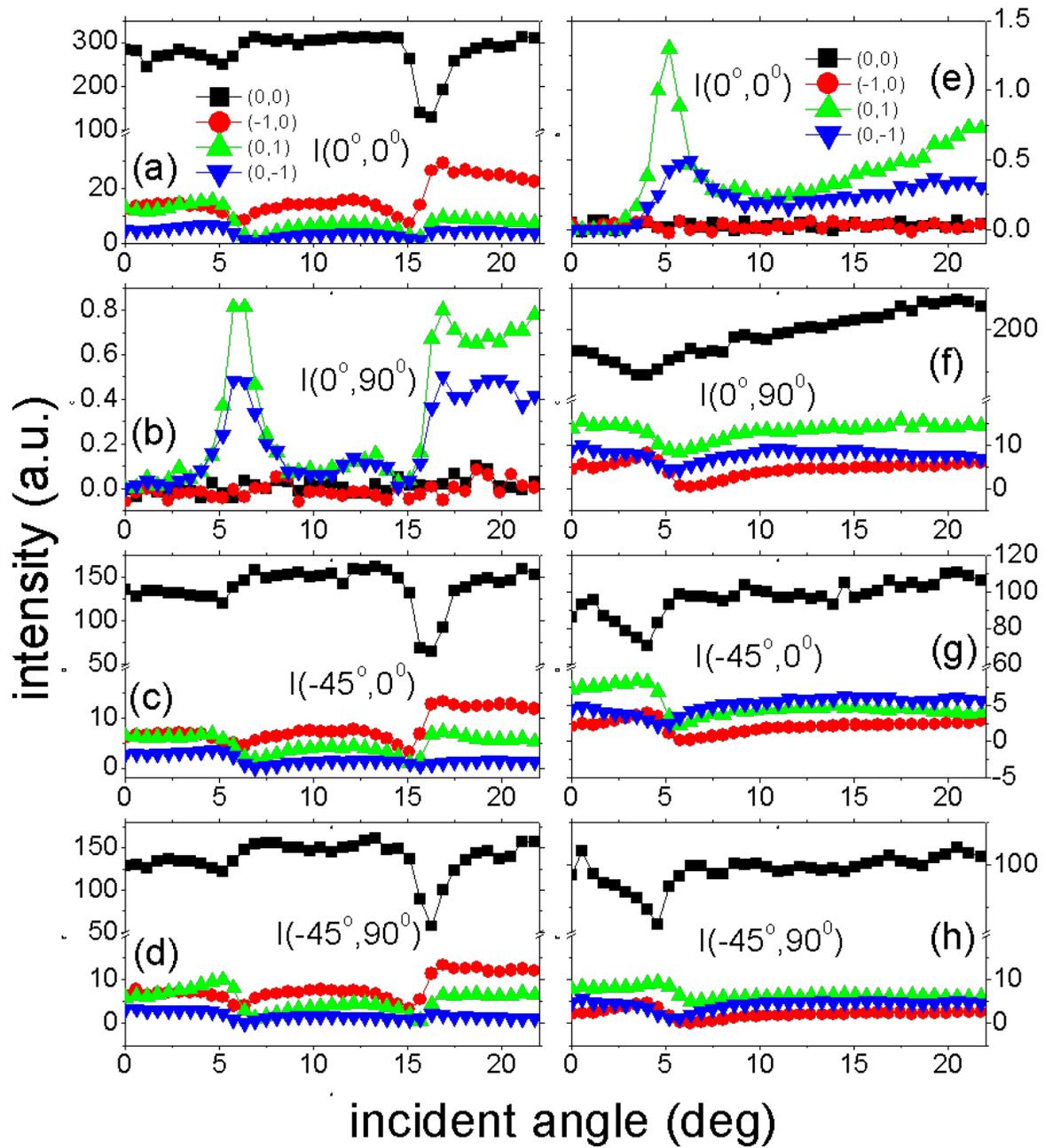

Fig. 1s. Plots of diffraction intensity of different orders as a function of incident angle. They are (a,e) $I(0°,0°)$, (b,f) $I(90°,0°)$, (c,g) $I(-45°,0°)$, and (d,h) $I(-45°,90°)$ of the {0,0}, {-1,0}, and {0,±1} diffraction orders as a function of incident angle under (a-d) p- and (e-h) s-excitations.

The following figure is the comparison between the angle-dependent p-polarized reflectivity taken from the goniometer and the angular intensity plot measured by the Fourier-space microscope at λ = 633 nm.

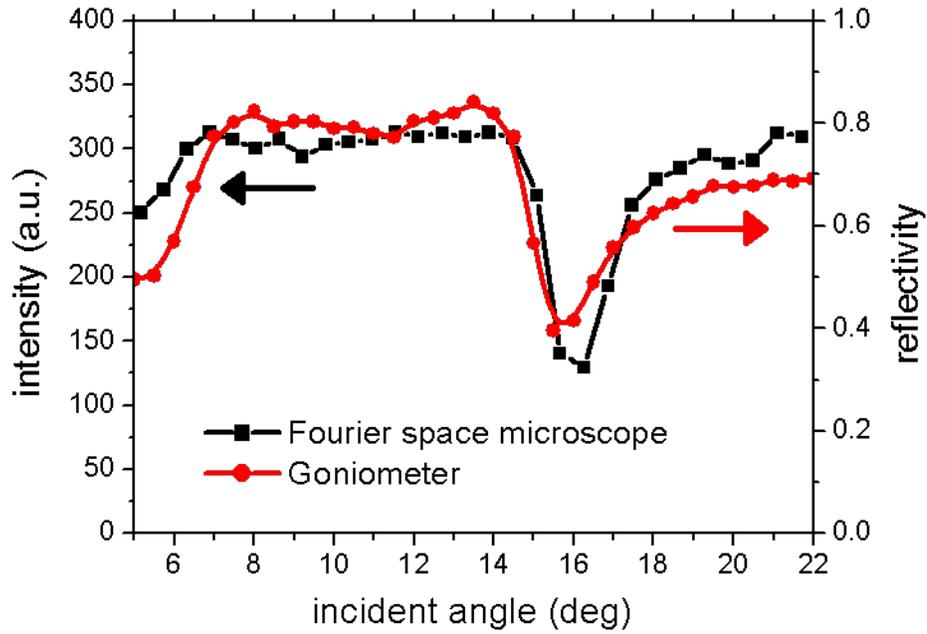

Fig. 2s. Plots of the angle-dependent p-polarized reflectivity taken from the goniometer and the angular intensity plot measured by the Fourier-space microscope at λ = 633 nm.

## II. The derivations of the discrete dipole model and the transition probabilities

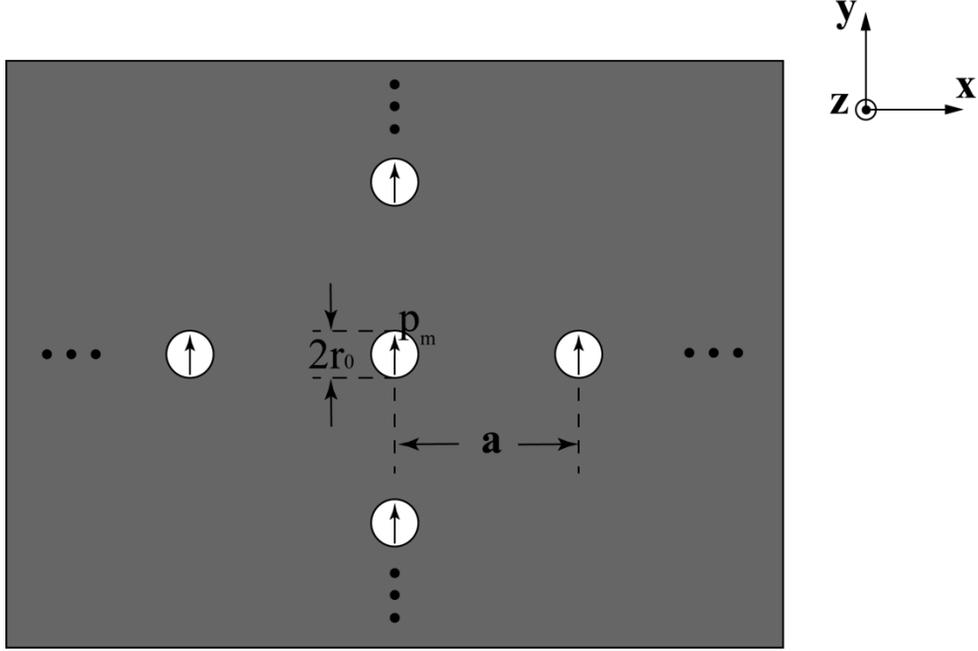

Fig. 2s. Schematic picture of 2D hole array under consideration.

As shown in the above figure, under the condition $r_0 \ll a$ and $kr_0 \ll 2\pi$, we model each hole as electric dipole $\mathbf{p}_{\vec{m}} = \vec{\vec{\alpha}}_{\vec{m}} \cdot \mathbf{E}_{\vec{m}}^{loc}$, in which $\vec{m} = (m_x, m_y)$ and $\mathbf{E}_{\vec{m}}^{loc}$ is the local electric field. Assume the external time-period driving electric filed is $\mathbf{E}_{\vec{m}}^{ext} e^{-i\omega t}$, $\mathbf{E}_{\vec{m}}^{loc}$ could then be expressed as:

$$\mathbf{E}_{\vec{m}}^{loc} = \mathbf{E}_{\vec{m}}^{ext} + \vec{\vec{W}}_{\vec{m}\vec{n}}(\mathbf{R}_{\vec{m}} - \mathbf{R}_{\vec{n}}) \cdot \mathbf{p}_{\vec{n}}, \tag{1.1}$$

where $\vec{\vec{W}}_{\vec{m}\vec{n}}(\mathbf{R}_{\vec{m}} - \mathbf{R}_{\vec{n}})$ is the dyadic Green's function defined as

$$\mathbf{W}_{\vec{m}\vec{n}}^{ij}(\mathbf{r}) = \begin{cases} \dfrac{k_0^3}{4\pi\varepsilon_0}\left[A(k_0 r)\delta_{ij} + B(k_0 r)\dfrac{r_i r_j}{r^2}\right], & \text{for } \vec{m} \neq \vec{n} \\ 0, & \text{for } \vec{m} = \vec{n} \end{cases}, \tag{1.2}$$

$$\begin{aligned} A(x) &= \left(x^{-1} + ix^{-2} - x^{-3}\right)e^{ix} \\ B(x) &= \left(-x^{-1} - 3ix^{-2} + 3x^{-3}\right)e^{ix} \end{aligned}, \tag{1.3}$$

with $\mathbf{r}_{\vec{m}\vec{n}} = \mathbf{R}_{\vec{m}} - \mathbf{R}_{\vec{n}} = (m_x - n_x)a\hat{x} + (m_y - n_y)a\hat{y}$, $k_0 = \omega/c_h$, $c_h$ is the speed of light in the background medium, and $i, j = 1, 2, 3$ stands for the component indices in Cartesian coordinates. Based on the above equations, we arrive at:

$$\mathbf{p}_{\vec{m}} = \vec{\vec{\alpha}}_{\vec{m}} \cdot \mathbf{E}_{\vec{m}}^{\text{ext}} + \vec{\vec{\alpha}}_{\vec{m}} \cdot \vec{\vec{W}}_{\vec{m}\vec{n}}(\mathbf{r}_{\vec{m}\vec{n}}) \cdot \mathbf{p}_{\vec{n}}$$
$$\Rightarrow \left[ \vec{\vec{\alpha}}_{\vec{m}}^{-1} \delta_{\vec{m}\vec{n}} - \vec{\vec{W}}_{\vec{m}\vec{n}}(\mathbf{r}_{\vec{m}\vec{n}}) \right] \cdot \mathbf{p}_{\vec{n}} = \mathbf{E}_{\vec{m}}^{\text{ext}} \quad (1.4)$$

In the case shown in Fig. 3s, the Bloch theorem should hold so that:

$$\mathbf{p}_{\vec{m}} = \tilde{\mathbf{p}} \, e^{i\vec{k} \cdot \vec{R}_{\vec{m}}}, \quad (1.5)$$

where $\vec{k}$ is the Bloch wavevector, and $\tilde{\mathbf{p}}$ corresponds to the dipole distribution. Substitute Eq.(1.5) into Eq.(1.4) and we have

$$\left[ \vec{\vec{\alpha}}_{\vec{m}}^{-1} \delta_{\vec{m}\vec{n}} - \vec{\vec{W}}_{\vec{m}\vec{n}}(\mathbf{r}_{\vec{m}\vec{n}}) \right] \cdot \tilde{\mathbf{p}} \, e^{i\vec{k} \cdot \vec{R}_{\vec{n}}} = \tilde{\mathbf{E}}^{\text{ext}} e^{i\vec{k} \cdot \vec{R}_{\vec{m}}}$$
$$\Rightarrow \left[ \vec{\vec{\alpha}}^{-1} - \sum_{\vec{n}} \vec{\vec{W}}_{\vec{m}\vec{n}}(\mathbf{r}_{\vec{m}\vec{n}}) \exp(-i\vec{k} \cdot \vec{r}_{\vec{m}\vec{n}}) \right] \cdot \tilde{\mathbf{p}} = \tilde{\mathbf{E}}^{\text{ext}}, \quad (1.6)$$

where $\mathbf{r}_{\vec{m}\vec{n}} = \mathbf{R}_{\vec{m}} - \mathbf{R}_{\vec{n}} = (m_x - n_x)a\hat{x} + (m_y - n_y)a\hat{y} \equiv \vec{n}a$. We introduce the following $3 \times 3$ matrix as:

$$\tilde{\mathbf{M}} = \vec{\vec{\alpha}}^{-1} - \sum_{\ell_1, \ell_2 = -\infty}^{+\infty} \vec{\vec{W}}(\vec{n}a) e^{-i\vec{k} \cdot \vec{n}a} \equiv \vec{\vec{\alpha}}^{-1} - \vec{\vec{S}}, \quad (1.7)$$

where the second term is the lattice sum $\vec{\vec{S}}$. Eq.(1.6) is then reduced to

$$\tilde{\mathbf{M}} \cdot \tilde{\mathbf{p}} = \tilde{\mathbf{E}}^{\text{ext}}. \quad (1.8)$$

It is worth to notice that $\tilde{\mathbf{M}}$'s dimension is $3 \times 3$ and we denote the eigenvalues and eigenvectors of $\tilde{\mathbf{M}}$ as $\{\lambda_\ell, \tilde{\mathbf{p}}_\ell; \ell = 1, 2, 3\}$. Define the eigen-polarizability as

$$\alpha_\ell^{\text{eig}} = \frac{1}{\lambda_\ell}, \quad (1.9)$$

and then we could use $\text{Im}(\alpha_\ell^{\text{eig}})/(\varepsilon_0 a^3)$ (the denominator will make the quantity dimensionless) to identify the resonance behavior of the system.

We could write the lattice sum as:

$$\ddot{\mathbf{S}} = \sum_{n_1,n_2=-\infty}^{+\infty} \ddot{\mathbf{W}}(\vec{n}a) e^{-i\vec{k}\cdot\vec{n}a}$$

$$= \sum_{\vec{n}\neq 0} \frac{k_0^3}{4\pi\varepsilon_0} \begin{pmatrix} A(k_0|\vec{n}|a) + B(k_0|\vec{n}|a)\frac{n_x^2}{n_x^2+n_y^2} & B(k_0|\vec{n}|a)\frac{n_x n_y}{n_x^2+n_y^2} & \\ B(k_0|\vec{n}|a)\frac{n_x n_y}{n_x^2+n_y^2} & A(k_0|\vec{n}|a) + B(k_0|\vec{n}|a)\frac{n_y^2}{n_x^2+n_y^2} & \\ & & A(k_0|\vec{n}|a) \end{pmatrix} e^{-i\vec{k}\cdot\vec{n}a}$$

$$\equiv \frac{1}{\varepsilon_0 a^3}\begin{pmatrix} S_T & S^{xy} & \\ S^{xy} & S_T & \\ & & S_L \end{pmatrix},$$

(1.10)

where $S_T, S_L, S^{xy}$ are dimensionless quantities that are determined by the lattice $a$. With this definition, Eq.(1.7) becomes

$$\tilde{\mathbf{M}} = \vec{\boldsymbol{\alpha}}^{-1} - \ddot{\mathbf{S}} = \begin{pmatrix} \alpha_\parallel^{-1} - \frac{1}{\varepsilon_0 a^3} S_T & -\frac{1}{\varepsilon_0 a^3} S^{xy} & \\ -\frac{1}{\varepsilon_0 a^3} S^{xy} & \alpha_\parallel^{-1} - \frac{1}{\varepsilon_0 a^3} S_T & \\ & & \alpha_\perp^{-1} - \frac{1}{\varepsilon_0 a^3} S_L \end{pmatrix}, \qquad (1.11)$$

where $\alpha_{\parallel,\perp}$ denote the polarizabilities of each hole. Diagonalization of $\tilde{\mathbf{M}}$ gives

$$\left(\alpha_1^{eig}\right)^{-1} = \lambda_1 = \alpha_\parallel^{-1} - \frac{1}{\varepsilon_0 a^3} S_T - \frac{1}{\varepsilon_0 a^3} S^{xy}, \quad \tilde{\mathbf{p}}_1 = \frac{1}{\sqrt{2}}(1,1,0)^T, \qquad (1.12)$$

$$\left(\alpha_2^{eig}\right)^{-1} = \lambda_2 = \alpha_\parallel^{-1} - \frac{1}{\varepsilon_0 a^3} S_T + \frac{1}{\varepsilon_0 a^3} S^{xy}, \quad \tilde{\mathbf{p}}_2 = \frac{1}{\sqrt{2}}(-1,1,0)^T, \qquad (1.13)$$

$$\left(\alpha_3^{eig}\right)^{-1} = \lambda_3 = \alpha_\perp^{-1} - \frac{1}{\varepsilon_0 a^3} S_L, \quad \tilde{\mathbf{p}}_3 = (0,0,1)^T. \qquad (1.14)$$

### III. Formulation by the Density Matrix Method

Since eigen-polarizations and eigen-vectors are ready, we use them to formulate the density matrix in this section. First of all, the formal solutions of the system could be written as

$$|\mathbf{p}^\ell(\omega,\mathbf{k})\rangle \equiv \frac{a}{2\pi}\begin{pmatrix}\vdots\\ e^{i\mathbf{k}\cdot\mathbf{R_m}}\tilde{\mathbf{p}}^\ell \\ \vdots\end{pmatrix}, \qquad (2.1)$$

where $|\mathbf{p}^\ell(\omega,\mathbf{k})\rangle$ has infinite dimension and $\mathbf{m}$ sorts in the way that $m_x$ runs first and $m_y$ comes the next. The orthogonality condition gives:

$$\langle\mathbf{p}^\ell(\omega,\mathbf{k})|\mathbf{p}^{\ell'}(\omega,\mathbf{k}')\rangle = \left(\frac{a}{2\pi}\right)^2 (\tilde{\mathbf{p}}^{\ell,H}\tilde{\mathbf{p}}^{\ell'}) \sum_{\mathbf{m}=-\infty}^{+\infty} e^{i(\mathbf{k}'-\mathbf{k})\cdot\mathbf{R_m}} = \delta_{\ell\ell'} \sum_{\mathbf{n}=-\infty}^{+\infty} \delta\left(\mathbf{k}-\mathbf{k}'+\mathbf{n}\frac{2\pi}{a}\right),$$

(2.2)

where $\ell = 1,2,3$. It is noted that the set $\{|\mathbf{p}^\ell(\omega,\mathbf{k})\rangle, \mathbf{k}\in 1\text{st Brillouin Zone}\}$ forms a complete set to describe the system. Similar to the density matrix operator, we could introduce the following operator as:

$$\hat{\rho}(\omega) = \int d\mathbf{k}\sum_\ell \alpha_\ell^{\text{eig}}(\omega,\mathbf{k})|\mathbf{p}^\ell(\omega,\mathbf{k})\rangle\langle\mathbf{p}^\ell(\omega,\mathbf{k})|, \qquad (2.3)$$

to model the response of system shown in Fig. 2s. If we assume the external EM wave possesses the form

$$\mathbf{E}^{in} = \sum_{\sigma=1}^{2} E_\sigma^{in}\hat{e}_\sigma(\mathbf{k}^{in})e^{i\mathbf{k}^{in}\cdot\mathbf{r}} \quad and \quad \mathbf{E}^{out} = \sum_{\sigma=1}^{2} E_\sigma^{out}\hat{e}_\sigma(\mathbf{k}^{out})e^{i\mathbf{k}^{out}\cdot\mathbf{r}}, \qquad (2.4)$$

where $\hat{e}_\sigma$ is the unit vector of the polarized plane waves, then

$$|\mathbf{E}^{in}\rangle \equiv \begin{pmatrix}\vdots\\ \sum_{\sigma=1}^{2} E_\sigma^{in}\hat{e}_\sigma e^{i\mathbf{k}^{in}\cdot\mathbf{R_m}} \\ \vdots\end{pmatrix}. \qquad (2.5)$$

With these definitions, we calculate the response function as

$$R_{\sigma',\sigma}(\omega,\mathbf{k}^{in},\mathbf{k}^{out}) = \langle \mathbf{E}^{out} | \hat{\rho} | \mathbf{E}^{in} \rangle$$

$$= \int d\mathbf{k} \sum_{\ell} \alpha_{\ell}^{eig}(\omega,k) \sum_{\mathbf{p},\mathbf{q}} E_{\sigma'}^{out} \hat{e}_{\sigma'}^{H} e^{-i\mathbf{k}^{out}\cdot\mathbf{R}_{\mathbf{p}}} \cdot \left(\frac{a}{2\pi}\right)^2 e^{i\mathbf{k}\cdot(\mathbf{R}_{\mathbf{p}}-\mathbf{R}_{\mathbf{q}})} (\tilde{\mathbf{p}}^{\ell}\tilde{\mathbf{p}}^{\ell,H}) \cdot E_{\sigma}^{in} \hat{e}_{\sigma} e^{i\mathbf{k}^{in}\cdot\mathbf{R}_{\mathbf{q}}}$$

$$= \int d\mathbf{k} \sum_{\ell} \alpha_{\ell}^{eig}(\omega,k) \sum_{\mathbf{p},\mathbf{q}} E_{\sigma}^{in} E_{\sigma'}^{out} \left(\frac{a}{2\pi}\right)^2 e^{-i\mathbf{k}^{out}\cdot\mathbf{R}_{\mathbf{p}}} e^{i\mathbf{k}\cdot(\mathbf{R}_{\mathbf{p}}-\mathbf{R}_{\mathbf{q}})} e^{i\mathbf{k}^{in}\cdot\mathbf{R}_{\mathbf{q}}} \left(\hat{e}_{\sigma'}^{H} \cdot \tilde{\mathbf{p}}^{\ell}\tilde{\mathbf{p}}^{\ell,H} \cdot \hat{e}_{\sigma}\right)$$

$$= \int d\mathbf{k} \sum_{\ell} \alpha_{\ell}^{eig}(\omega,k) \sum_{\mathbf{s},\mathbf{t}} E_{\sigma}^{in} E_{\sigma'}^{out} \left(\frac{2\pi}{a}\right)^2 \delta\left(\mathbf{k}-\mathbf{k}^{out}+\mathbf{s}\frac{2\pi}{a}\right) \delta\left(\mathbf{k}^{in}-\mathbf{k}+\mathbf{t}\frac{2\pi}{a}\right) \Pi_{\sigma',\sigma}^{\ell}(\omega,\mathbf{k},\mathbf{k}^{in},\mathbf{k}^{out})$$

$$= \sum_{\ell}\sum_{\mathbf{s},\mathbf{t}} \alpha_{\ell}^{eig}\left(\omega,\mathbf{k}^{in}+\mathbf{t}_0\frac{2\pi}{a}\right) E_{\sigma}^{in} E_{\sigma'}^{out} \left(\frac{2\pi}{a}\right)^2 \delta\left(\mathbf{k}^{in}-\mathbf{k}^{out}+(\mathbf{s}+\mathbf{t}_0)\frac{2\pi}{a}\right) \Pi_{\sigma',\sigma}^{\ell}\left(\omega,\mathbf{k}^{in}+\mathbf{t}_0\frac{2\pi}{a},\mathbf{k}^{in},\mathbf{k}^{out}\right)$$

$$\equiv \sum_{\mathbf{s}} \left(\frac{2\pi}{a}\right)^2 f_{\sigma',\sigma}(\omega_0,\mathbf{k}^{in},\mathbf{k}^{out}) \delta\left(\mathbf{k}^{in}-\mathbf{k}^{out}+(\mathbf{s}+\mathbf{t}_0)\frac{2\pi}{a}\right)$$

(2.6)

where $\mathbf{t}_0$ is chosen so that $\mathbf{k}^{in}+\mathbf{t}_0\frac{2\pi}{a}$ lies in the first Brillouin zone, and

$$f_{\sigma',\sigma}(\omega,\mathbf{k}^{in},\mathbf{k}^{out}) = \sum_{\ell} E_{\sigma'}^{out} E_{\sigma}^{in} \alpha_{\ell}^{eig}\left(\omega,\mathbf{k}^{in}+\mathbf{t}_0\frac{2\pi}{a}\right) \Pi_{\sigma',\sigma}^{\ell}\left(\omega,\mathbf{k}^{in}+\mathbf{t}_0\frac{2\pi}{a},\mathbf{k}^{in},\mathbf{k}^{out}\right), \quad (2.7)$$

$$\Pi_{\sigma',\sigma}^{\ell}(\omega,k,\mathbf{k}^{in},\mathbf{k}^{out}) = \hat{e}_{\sigma'}^{H}(\mathbf{k}^{out}) \cdot \tilde{\mathbf{p}}^{\ell}\tilde{\mathbf{p}}^{\ell,H} \cdot \hat{e}_{\sigma}(\mathbf{k}^{in}). \quad (2.8)$$

It is worth to notice that the response of the system behaviors like reflection gratings so that the Bragg scattering yields the condition of momentum conversation and the intra unit cell properties give the transition amplitude. Next we introduce the transition probability from one eigen-state $|\mathbf{E}^{in}\rangle$ to another eigen-state $|\mathbf{E}^{out}\rangle$ such that:

$$P_{\sigma',\sigma}(\omega,\mathbf{k}^{in},\mathbf{k}^{out}) \equiv \frac{|f_{\sigma',\sigma}(\omega,\mathbf{k}^{in},\mathbf{k}^{out})|^2}{I^{out}*I^{in}}, \quad (2.9)$$

where $I^{in} = \frac{1}{2}\varepsilon_0|E^{in}|^2 V$, and $I^{out} = \frac{1}{2}\varepsilon_0|E^{out}|^2 V$. Noted that $P_{\sigma',\sigma}$ is a dimensionless quantity that reflects the transition probability between two eigen-states in free space. This transition probability can be mathematically evaluated as:

$$P_{\sigma',\sigma}\left(\omega,\mathbf{k}^{in},\mathbf{k}^{out}\right)=\left|\sum_{\ell}\frac{2}{\varepsilon_0 a^3}\alpha_\ell^{eig}\left(\omega,\mathbf{k}^{in}+\mathbf{t}_0\frac{2\pi}{a}\right)\Pi_{\sigma',\sigma}^{\ell}\left(\omega,\mathbf{k}^{in}+\mathbf{t}_0\frac{2\pi}{a},\mathbf{k}^{in},\mathbf{k}^{out}\right)\right|^2$$

$$=\left|\langle\hat{e}_{\sigma'}\left(\mathbf{k}^{out}\right)|\left[\sum_{\ell}\frac{2}{\varepsilon_0 a^3}\alpha_\ell^{eig}\left(\omega,\mathbf{k}^{in}+\mathbf{t}_0\frac{2\pi}{a}\right)|\tilde{\mathbf{p}}^\ell\rangle\langle\tilde{\mathbf{p}}^\ell|\right]|\hat{e}_\sigma\left(\mathbf{k}^{in}\right)\rangle\right|^2$$

$$\equiv\left|\langle\hat{e}_{\sigma'}\left(\mathbf{k}^{out}\right)|\mathbf{T}|\hat{e}_\sigma\left(\mathbf{k}^{in}\right)\rangle\right|^2$$

(2.10)

In our special case, **T** matrix is

$$\mathbf{T}=\frac{2}{\varepsilon_0 a^3}\begin{pmatrix}\frac{\alpha_\parallel^{-1}-\frac{1}{\varepsilon_0 a^3}S_T}{\left(\alpha_\parallel^{-1}-\frac{1}{\varepsilon_0 a^3}S_T\right)^2-\left(\frac{1}{\varepsilon_0 a^3}S^{xy}\right)^2} & \frac{\frac{1}{\varepsilon_0 a^3}S^{xy}}{\left(\alpha_\parallel^{-1}-\frac{1}{\varepsilon_0 a^3}S_T\right)^2-\left(\frac{1}{\varepsilon_0 a^3}S^{xy}\right)^2} & \\ \frac{\frac{1}{\varepsilon_0 a^3}S^{xy}}{\left(\alpha_\parallel^{-1}-\frac{1}{\varepsilon_0 a^3}S_T\right)^2-\left(\frac{1}{\varepsilon_0 a^3}S^{xy}\right)^2} & \frac{\alpha_\parallel^{-1}-\frac{1}{\varepsilon_0 a^3}S_T}{\left(\alpha_\parallel^{-1}-\frac{1}{\varepsilon_0 a^3}S_T\right)^2-\left(\frac{1}{\varepsilon_0 a^3}S^{xy}\right)^2} & \\ & & \frac{1}{\alpha_\perp^{-1}-\frac{1}{\varepsilon_0 a^3}S_L}\end{pmatrix},$$

(2.11)

yielding the transition matrix used in the main text.